\documentstyle[12pt]{article}
\begin{document}

\baselineskip 6mm
\renewcommand{\thefootnote}{\fnsymbol{footnote}}

\newcommand{\nc}{\newcommand}
\newcommand{\rnc}{\renewcommand}

%\headheight=0truein
%\headsep=0truein
%\topmargin=0truein
%\oddsidemargin=0truein
%\evensidemargin=0truein
%\textheight=9truein
%\textwidth=6.5truein

\rnc{\baselinestretch}{1.44}    % 1.5 spacing btwn text lines
\setlength{\jot}{7pt}       % spacing btwn the rows of an eqnarray
\rnc{\arraystretch}{1.64}   % spacing btwn the rows of a non-eqn array

%%%%%%%%%%%%%%%%%%%%%%%%%%%%%%%%%%%%%%%%%%%%%%%%%%%%%%%%%%%%%%%%%
%                                                               %
%                NEW COMMANDS AND MACROS                        %
%                                                               %
%%%%%%%%%%%%%%%%%%%%%%%%%%%%%%%%%%%%%%%%%%%%%%%%%%%%%%%%%%%%%%%%%

%%%%% Simplify some frequently used LaTeX commands %%%%%

\nc{\be}{\begin{equation}}

\nc{\ee}{\end{equation}}

\nc{\bea}{\begin{eqnarray}}

\nc{\eea}{\end{eqnarray}}

\nc{\ben}{\begin{eqnarray*}}

\nc{\een}{\end{eqnarray*}}

\nc{\xx}{\nonumber\\}

\nc{\ct}{\cite}

\nc{\la}{\label}

\nc{\eq}[1]{(\ref{#1})}

\nc{\newcaption}[1]{\centerline{\parbox{6in}{\caption{#1}}}}

\nc{\fig}[3]{

\begin{figure}
\centerline{\epsfxsize=#1\epsfbox{#2.eps}}
\newcaption{#3. \label{#2}}
\end{figure}
}

%%% Double line letters %%%

\def\IR{{\hbox{{\rm I}\kern-.2em\hbox{\rm R}}}}
\def\IB{{\hbox{{\rm I}\kern-.2em\hbox{\rm B}}}}
\def\IN{{\hbox{{\rm I}\kern-.2em\hbox{\rm N}}}}
\def\IC{\,\,{\hbox{{\rm I}\kern-.59em\hbox{\bf C}}}}
\def\IZ{{\hbox{{\rm Z}\kern-.4em\hbox{\rm Z}}}}
\def\IP{{\hbox{{\rm I}\kern-.2em\hbox{\rm P}}}}
\def\IH{{\hbox{{\rm I}\kern-.4em\hbox{\rm H}}}}
\def\ID{{\hbox{{\rm I}\kern-.2em\hbox{\rm D}}}}

%%%%% Roman pont in math

\def\Tr{{\rm Tr}\,}
\def\det{{\rm det}}
\def\sv{{\cal v}}

%%%%% Special Letters

\def\vare{\varepsilon}
\def\barz{\bar{z}}
\def\barw{\bar{w}}

\begin{titlepage}
%---------------- preprint number ---------------
%\hfill\parbox{4cm}
\begin{flushright}
{IP/BBSR/2004-28}\\
hep-th/0503116
\end{flushright}

\vspace{25mm}
\begin{center}
%------------------------ title ------------------------
{\Large{\bf ${\cal N}= 2$ $\sigma$-model Action 
on Non(anti)commutative \\ 
\vskip 0.2cm 
Superspace} } 

\vspace{12mm}
%---------------- authors and addresses ----------------
B.~ Chandrasekhar\footnote{chandra@iopb.res.in} 
\\[3mm]
{ Institute of Physics, Bhubaneswar 751 005, INDIA} \\
\end{center}

\thispagestyle{empty}

\vskip 2cm

%----------------------- abstract ----------------------

\centerline{\bf ABSTRACT}
\vskip 4mm
\noindent

We show that the infinite series in the classical
action for non(anti)commutative ${\cal N}=2,\:\sigma$-models
in two dimensions, can be resummed by using constraint equations 
of the auxiliary fields. We argue that the resulting action takes
a standard form and the target space is necessarily smeared by
terms dependent on the deformation parameter.

\vspace{2cm}

\today

\end{titlepage}

\newpage

Nonlinear Sigma models in two dimensions on general 
target spaces have given inumerable physical insights
in to various branches of Physics and Mathematics. 
Supersymmetric extension of bosonic sigma models unfolded new avenues
in the study of complex geometry. In particular, extension to
${\cal N}=2$ supersymmetry forces the target manifold to be K\"{a}hler 
%and ${\cal N}=4$ supersymmetry requires the manifold to be hyperK\"{a}hler
\cite{Zumino:1979et}-\cite{Alvarez-Gaume:hn}, which 
plays a crucial role in showing the consistency of these
sigma models at the quantum level. The importance of 
the sigma model approach to the study of string compactifications 
on Ricci-flat K\"{a}hler manifolds (otherwise Calabi-yau manifolds) 
and also string propagation in arbitrary background fields is well known.
%Infact, the equations of motion of the background fields follow 
%from conformal invariance of the underlying sigma model. 
Due to such varied interests, Non-commutative~\cite{Seiberg:1999vs}
deformations of these sigma models have also been actively 
pursued~\cite{Girotti:2001gs}. 

Recently, the emergence of a new non(anti)commutative 
superspace deformation has opened up an interesting arena, 
ensuing from Dijkgraaf-Vafa relations and the consideration of
superstrings in graviphoton background 
fields~\cite{Ooguri:2003qp,Seiberg:2003yz,Berkovits:2003kj}. 
However, it turns out that such a superspace deformation is only
possible in a Euclidean space~\cite{Klemm:2001yu}.
Following the work in~\cite{Seiberg:2003yz}, various aspects of 
supersymmetric theories with non(anti)commutative deformation are being 
actively investigated. Here, one only retains half of the
supersymmetry generators of the theory. For other kinds of supersymmetric
deformations and harmonic superspace approaches  
see~\cite{Ferrara:2000mm,Ivanov:2003te}.

To understand the conformal structure of non(anti)commutative theories
in two dimensions, we studied the classical aspects of   
$\sigma$-models defined on non(anti)commutative (NAC)
superspace with a general K\"{a}hler 
potential~\cite{Chandrasekhar:2003uq,Chandrasekhar:2004ti}. 
It was shown that due to the NAC deformation, 
the classical action has infinitely many terms. Despite this fact, it 
is possible to write down the action in a closed form, after
identifying the emergence of a series expansion in the 
NAC parameter. However, in its present form, the action appears
complicated and is inaccessible even in the classical domain. Moreover,
the auxiliary fields cannot be eliminated directly, due to the infinite
series. On the otherhand, to prove the consistency of these  NAC
$\sigma$-models, it is important
to establish various results, such as quantum renormalizability, 
conformal invariance etc., Conformal invariance for instance,
is essential to obtain the equations of motion of superstrings propagating
in arbitrary background fields. 

To understand such interesting aspects, in this letter, we show that
the infinite series in the classical action of the NAC sigma models
can be resummed and in the resulting action the target space appears
smeared. We argue that the new form of the action establishes a 
good connection with the $C=0$ $\sigma$-models and may turn out be useful in 
drawing important conclusions about the quantum and conformal structure.

Note added: During the final stages of this draft, 
ref.~\cite{Alvarez-Gaume:2005pj} appeared on the arXiv, where the 
resummation of the action with superpotential terms is discussed
and the smearing of the target space is understood as deformation 
of Zumino's lagrangian.

We follow the notations of~\cite{Chandrasekhar:2004ti}. The 2D superspace
variables are $\theta^{\pm},{\bar \theta}^{\pm},x^0,x^3 $ and $y^0,y^3$ stand
for the chiral coordinates. The world-sheet is deformed as:
\be \label{Cdeformation}
\{\theta^{\alpha},\theta^{\beta}\}= C^{\alpha\beta},
\ee
where $\alpha,\beta$ run over $+,-$. Rest of the (anti)commutation
relations between $\theta,{\bar \theta},x^0,x^3$ are 
zero~\cite{Seiberg:2003yz,Chandrasekhar:2004ti}. 
The ${\cal N}=2$ supersymmetry generators 
$Q_{\pm} = -\frac{\partial}{\partial\theta^{\pm}}, ~ 
\bar{Q}_{\pm} = -\frac{\partial}{\partial{\bar\theta}^{\pm}} 
- 2 i \theta^{\pm}
\left(\frac{\partial}{\partial y^0}
\pm \frac{\partial}{\partial y^3} \right)$ satisfy,
\be \label{q}
\{Q_{\pm}, {\bar Q}_{\pm} \} = -2i (\frac{\partial}{\partial y^0}
\pm \frac{\partial}{\partial y^3}),
\ee
but, $\{{\bar Q}_{\pm}, {\bar Q}_{\pm} \} \neq 0$, which is due to
the nonlinearity of $\bar Q$'s. As a result we only work with the 
${\cal N}=1/2$ supersymmetry generators, $Q_{\pm}$. Here, we also
note that the superspace deformation in eqn. (\ref{Cdeformation}),
is only possible in Euclidean space and there are no complex 
conjugation conditions 
on any fields~\cite{Lukierski:1986jw,TERA,Chandrasekhar:2004ti}.

The component form of the ${\cal N}=2$ $\sigma$-model action with a 
superspace deformation as given in eqn. (\ref{Cdeformation}), 
can be derived from a superspace integral over the K\"{a}hler potential.
Details of the derivation are given 
in~\cite{Chandrasekhar:2003uq,Chandrasekhar:2004ti} and below we only
quote the final result:
\bea \label{I}
%%% An overall factor of -4 has been taken out
I &=& -\sum_{n= 1}^{\infty}(-1)^n
(\det\,C)^{n-1}\int d^2x 
\frac{F^{i_1}F^{i_2}\cdots F^{i_{2n-2}} }{(2n-1)!}
\Bigl[\psi^{i_{2n-1}}_-
\psi^{i_{2n}}_+ {\bar F}^j \,
 {\mathcal K}_{,i_1i_2\cdots i_{2n}\bar j }
+\,\psi^{i_{2n-1}}_-{\bar\psi}^j_-
\psi^{i_2n}_+{\bar\psi}^k_+ \, \xx
&\times&{\mathcal K}_{,i_1i_2\cdots i_{2n} \bar j \bar k}
\:\Bigr] 
~+~ \sum_{n=0}^{\infty}
(\det\,C)^n (-1)^n %% 
\:\int d^2x \: F^{i_1}F^{i_2}\cdots F^{i_{2n-1}}  
\Bigl[ \,\frac{1}{(2n)!} F^{i_{2n}} \partial_{\xi^-} 
\partial_{\zeta^-}{\bar \phi}^j\,
 {\mathcal K}_{,i_1i_2\cdots i_{2n}\bar j }   \xx
&\!\!\!\!+\!\!\!\!&\!\!\!\!\!\!\!\!
\frac {1}{(2n+1)!} 
\Bigl(\,2n\,\psi^{i_{2n}}_-
\psi^{i_{2n+1}}_+\partial_{\xi^-} 
\partial_{\zeta^-}{\bar \phi}^j 
+ i F^{i_{2n}}\psi^{i_{2n+1}}_-
\partial_{\zeta^-}{\bar\psi}^j_-  
+ iF^{i_{2n}}\psi^{i_{2n+1}}_+
\partial_{\xi^-} {\bar \psi}_+ 
+ F^{i_{2n}}F^{i_{2n+1}}{\bar F}^j\,\Bigr)  \xx
&\times&{\mathcal K}_{,i_1i_2\cdots i_{2n+1} \bar j}
+ \frac{1}{(2n)!}\,F^{i_{2n}} 
\partial_{\xi^-}{\bar \phi}^j\partial_{\zeta^-}{\bar \phi}^k\,
{\mathcal K}_{,i_1i_2\cdots i_{2n} \bar j \bar k}
+\frac{1}{(2n+1)!}\,
\Bigl(\,2n
\psi^{i_{2n}}_-
\psi^{i_{2n+1}}_+ \partial_{\xi^-}
{\bar \phi}^j \partial_{\zeta^-}{\bar \phi}^k \xx 
&-& F^{i_{2n}}F^{i_{2n+1}}{\bar\psi}^j_-{\bar\psi}^k_+ 
- F^{i_{2n}} {\bar\psi}^j_-\psi^{i_{2n+1}}_-
 \partial_{\zeta^-}{\bar \phi}^k 
- F^{i_{2n}} {\bar\psi}^j_+\psi^{i_{2n+1}}_+ 
\partial_{\xi^-}{\bar \phi}^k 
\,\Bigr) {\mathcal K}_{,i_1i_2\cdots i_{2n+1} 
\bar j \bar k}  \:\Bigr] \,.
\eea
where $\Phi^i$ and $\bar{\Phi}^i$ are the ${\cal N}=2$ superfields 
and the scalar function ${\mathcal K}$ is the K\"{a}hler potential.
Further, we have for convenience,
$ \xi^- = \frac{1}{2}\,(x^0 - x^3) - i \theta^- {\bar \theta}^- ,\:
\zeta^- = \frac{1}{2}\,(x^0 + x^3) - i \theta^+ {\bar \theta}^+$.
We also have the notation:
\be \label{kahler}
{\mathcal K}_{,i_1i_2\cdots i_n{\bar j}_1
{\bar j}_2\cdots {\bar j}_m} =
{ \frac{\partial^{(n+m)}{\mathcal K}}
{\partial \Phi^{i_1}\partial \Phi^{i_2}\cdots\partial \Phi^{i_n}
\partial {\bar \Phi}^{j_1}\partial {\bar \Phi}^{j_2}\cdots
\partial {\bar \Phi}^{j_m} } }|_{\Phi^i= \phi^i,\bar{\Phi}^i= {\bar \phi}^i},
\ee
for the derivatives of the K\"{a}hler potential with respect to the
chiral and anti-chiral superfields evaluated at 
$\Phi^i= \phi^i$ and $\bar{\Phi}^i= {\bar \phi}^i$.
The above action for the non(anti)commutative
$\sigma$-models is a series expansion
in $(\det\,C)$. Despite the presence of infinite terms, the 
action  (\ref{I}) has been shown be invariant under the
${\cal N}=1/2$ supersymmetry of the 
theory~\cite{Chandrasekhar:2003uq}, which is a further check
that the action in the component form is correct.

For the special case of $C=0$, the action in eqn. (\ref{I}) reduces
to the standard ${\cal N}=2 \:\sigma$-model action. In that case, it
is possible to eliminate the auxiliary fields by 
their equations of motion, given as:
\be \label{F}
F^i = -\psi^{j}_- \psi^{k}_+ \, \Gamma^i_{jk}, 
\ee 
with a similar relation for  ${\bar F}$. The resulting non-linear 
$\sigma$ model action takes the form:
\bea \label{Io}
I_o &=& \int d^2x \,\Bigl[\,(\partial_{\xi^-}\phi^i
\partial_{\zeta^-}{\bar \phi}^j
+i\psi^i_- {\bf D}_{\zeta^-}{\bar\psi}^j_-
+ i\psi^i_+ {\bf D}_{\xi^-}
{\bar\psi}^j_+)
\:g_{i\bar j} - \psi^i_-{\bar\psi}^k_- 
\psi^j_+{\bar\psi}^l_+  {\mathcal R}_{j\bar k i \bar l}
\,\Bigr],
\eea 
with the covariant derivative defined as 
${\bf D}_{\xi^-} {\bar\psi}^i_+
= \partial_{\xi^-}{\bar\psi}^i_+ 
+  \Gamma^{\bar i}_{{\bar j}\bar k} {\bar\psi}^j_+ \,
\partial_{\xi^-}{\bar \phi}^k $ and similarly for 
${\bar\psi}^i_-$. Here, 
$g_{i\bar{j}}= \partial_i\partial_{\bar j}{\mathcal K}$ 
corresponds to the background metric, $\Gamma^i_{jk}$
denote the Christoffel symbols and 
${\mathcal R}_{j\bar k i \bar l}$ is the Riemann curvature
tensor of the target manifold. However, due to the infinite
number of terms in the $C\neq 0$ sigma model action (\ref{I}), 
solving for the auxiliary fields seems 
difficult~\cite{Chandrasekhar:2004ti}.
Here, we would first like to show that  $F$
still satisfies a non-trivial constraint equation, although this is not
obvious from the $n^{th}$ order action in (\ref{I}). 

It is possible to write down terms in the action to a particular order in 
$(\det\,C)$. Thus $I = I_0 + I_C$, to order  $(\det\,C)$ can be written as:
\bea \label{cone}
I &=& \int d^2x \Bigl[(\frac{1}{2}\partial_{\xi^-}\phi^i
\partial_{\zeta^-}{\bar \phi}^j
+\frac{1}{2}\partial_{\zeta^-}\phi^i
\partial_{\xi^-}{\bar \phi}^j
+i\psi^i_-\partial_{\zeta^-}{\bar\psi}^j_-
+ i\psi^i_+\partial_{\xi^-}
{\bar\psi}^j_+ + F^i{\bar F}^j)
g_{i\bar j} + \psi^i_- \psi^j_+{\bar F}^k \xx
&\times&\Gamma_{ij\bar k} 
+ (i \psi^i_-{\bar\psi}^j_-\partial_{\zeta^-}{\bar \phi}^k
+ i \psi^i_+{\bar\psi}^j_+
\partial_{\xi^-}{\bar \phi}^k  
- F^i{\bar\psi}^j_-{\bar\psi}^k_+) \Gamma_{i\bar j \bar k}
+ (\psi^i_-{\bar\psi}^k_- 
\psi^j_+{\bar\psi}^l_+ ) \partial_i \Gamma_{j\bar k \bar l} 
+ (\det\,C)\:
F^{p} \xx
&\times& \Bigl[ -\frac{F^{q}}{6}\Bigl\{
\psi^{l}_-\psi^{m}_+ {\bar F}^j 
\partial_p  \partial_q \Gamma_{lm\bar j } 
+\psi^{l}_-{\bar\psi}^j_-
\psi^{m}_+{\bar\psi}^k_+  
\partial_p  \partial_q \partial_l\Gamma_{m \bar j \bar k}
\Bigr\} -\frac{1}{2} F^{q} \partial_{\xi^-} 
\partial_{\zeta^-}{\bar \phi}^j \Gamma_{pq\bar j }  
-\frac {1}{6} 
\Bigl(2\psi^{q}_- \psi^{l}_+ \xx
&\times&
\partial_{\xi^-} 
\partial_{\zeta^-}{\bar \phi}^j 
+ i F^{q}\psi^{l}_-
\partial_{\zeta^-}{\bar\psi}^j_-  
+ iF^{q}\psi^{l}_+
\partial_{\xi^-}{\bar \psi}_+  
+ F^{q}F^{l}{\bar F}^j\,\Bigr) 
\partial_p \Gamma_{ql \bar j}         
- \frac{1}{2} \, F^{q} 
\partial_{\xi^-}{\bar \phi}^j\partial_{\zeta^-}{\bar \phi}^k\, 
\partial_p\Gamma_{q\bar j \bar k}  \xx
&-& \frac{1}{6}\,\Bigl(\,2 \psi^{q}_-
\psi^{l}_+ \partial_{\xi^-}
{\bar \phi}^j \partial_{\zeta^-}{\bar \phi}^k 
- F^{q}F^{l}{\bar\psi}^j_-{\bar\psi}^k_+ 
- F^{q} {\bar\psi}^j_-\psi^{l}_-
 \partial_{\zeta^-}{\bar \phi}^k 
- F^{q} {\bar\psi}^j_+\psi^{l}_+ 
\partial_{\xi^-}{\bar \phi}^k 
\,\Bigr) \partial_p  \partial_q \Gamma_{l 
\bar j \bar k}  \:\Bigr] \,.
\eea  
For lucidity, we confine the discussion to terms in
the action to first order in $(\det\,C)$ as given in eqn. (\ref{cone}). 
Generalization to all orders
can be easily done. After some rearrangement, the action takes the form: 
\bea \label{Ioc1}
I &=& \int d^2x \,\Bigl[\,(\frac{1}{2}\partial_{\xi^-}\phi^i
\partial_{\zeta^-}{\bar \phi}^j
+\frac{1}{2}\partial_{\zeta^-}\phi^i
\partial_{\xi^-}{\bar \phi}^j ) g_{i\bar j}
+  (\det\,C)\: F^{p} \,\{
\frac{1}{2} F^{q} \partial_{\xi^-} 
\partial_{\zeta^-}{\bar \phi}^j\,\Gamma_{pq\bar j } \xx  
&+& \frac {1}{3}\,\psi^{q}_- \psi^{l}_+ 
\partial_{\xi^-} 
\partial_{\zeta^-}{\bar \phi}^j  \partial_p\Gamma_{q l \bar j} 
+ \frac{1}{2} \, F^{q} 
\partial_{\xi^-}{\bar \phi}^j\partial_{\zeta^-}{\bar \phi}^k\, 
\partial_p\Gamma_{q\bar j \bar k} 
+ \frac{1}{3}\, \psi^{q}_-
\psi^{l}_+ \partial_{\xi^-}
{\bar \phi}^j \partial_{\zeta^-}{\bar \phi}^k \partial_p\partial_q 
\Gamma_{l\bar j \bar k} 
\} \xx
&+& \Bigl(i\psi^i_-\partial_{\zeta^-}{\bar\psi}^j_- 
+  i\psi^i_+\partial_{\xi^-}
{\bar\psi}^j_+ + F^i{\bar F}^j \Bigr)
{\tilde g}_{i\bar j} 
+ \psi^i_- \psi^j_+{\bar F}^k 
{\tilde \Gamma}_{ij\bar k} 
+ (i \psi^i_-{\bar\psi}^j_-\partial_{\zeta^-}{\bar \phi}^k
+ i \psi^i_+{\bar\psi}^j_+
\partial_{\xi^-}{\bar \phi}^k   \xx
&-&  F^i{\bar\psi}^j_-{\bar\psi}^k_+) 
{\tilde \Gamma}_{i\bar j \bar k} + (\psi^i_-{\bar\psi}^k_- 
\psi^j_+{\bar\psi}^l_+ ) \partial_i {\tilde \Gamma}_{j\bar k \bar l}
\,\Bigr]. 
\eea
%Here, we have used the notation 
%$\Gamma_{ij \bar k} = g_{l\bar k}\Gamma^l_{ij} $ and 
Here, new geometric
quantities with an additional tilde as seen in eqn. (\ref{Ioc1}), 
are redefined. For instance, the metric is redefined as:
\be \label{NewgG}
{\tilde g}_{i{\bar j}}\, = g_{i{\bar j}} - 
 \frac{1}{6}(\det\,C) F^{p}F^{q}  
\partial_p \partial_q  g_{i{\bar j}}.
\ee
and for the Christoffel symbols in the action (\ref{Ioc1}), we
have the new definition:
%have the new definition, 
%${\tilde \Gamma}^i_{kl} = {\tilde g}^{i{\bar j}}\, 
%{\tilde \Gamma}_{kl\bar j} $ where:
\be \label{NewGamma}
{\tilde \Gamma}_{kl{\bar j}}\, = \Gamma_{kl{\bar j}} - 
 \frac{1}{6}(\det\,C) F^{p}F^{q}  
\partial_p \partial_q  \Gamma_{kl{\bar j}}.
\ee

At this stage, the results in eqns. (\ref{NewgG}) and (\ref{NewGamma})
follow from the rearrangement done in the action (\ref{Ioc1}). 
However, it is not clear whether one can independently calculate the 
Christoffels from the redefined metric given
in  eqn. (\ref{NewgG}) and show that it is identical to the result
in eqn. (\ref{NewGamma}). We comment more on this issue later on.
This check is infact important, as it elucidates whether
the $C$-deformation on the worldsheet induces
any kind of torsion terms on the target space manifold.  

In the action (\ref{Ioc1}), most of the 
terms proportional to $(\det\,C)$ have been rearranged in such a way
that, they only modify the geometric quantities of the 
zeroth order action. However, some terms involving the derivatives
of bosonic scalar fields are left untouched. These terms can be 
rearranged by doing partial integrations as shown below:
\be \label{partial}
\partial_{\xi^-} 
\partial_{\zeta^-}{\bar \phi}^j\,\Gamma_{pq\bar j }
=  - \partial_{\xi^-}\phi^l
\partial_{\zeta^-}{\bar \phi}^j \partial_l\Gamma_{pq\bar j } 
-  \partial_{\xi^-}{\bar \phi}^l \partial_{\zeta^-}{\bar \phi}^j 
\partial_{\bar l}\Gamma_{pq\bar j }
\ee
Doing similar partial integration on other terms, we arrive at:
\bea \label{I0new}
I &=& \int d^2x \,\Bigl[\,
(\frac{1}{2}\partial_{\xi^-}\phi^i
\partial_{\zeta^-}{\bar \phi}^j
+\frac{1}{2}\partial_{\zeta^-}\phi^i
\partial_{\xi^-}{\bar \phi}^j )\, (
g_{i\bar j} - (\det\,C)\,F^p \{ 
\frac{1}{2} F^{q}\, \partial_i\Gamma_{pq\bar j } 
+ \frac{1}{3}\, \psi^{q}_-
\psi^{l}_+ \,\partial_i\partial_q 
\Gamma_{l\bar j \bar k} 
\} \, ) \xx
&+& (i\psi^i_-\partial_{\zeta^-}{\bar\psi}^j_-
+ i\psi^i_+\partial_{\xi^-}
{\bar\psi}^j_+ + F^i{\bar F}^j )
{\tilde g}_{i\bar j} 
+ (\psi^i_-
\psi^j_+{\bar F}^k) 
{\tilde \Gamma}_{ij\bar k} \xx
&+& (i \psi^i_-{\bar\psi}^j_-\partial_{\zeta^-}{\bar \phi}^k
+ i \psi^i_+{\bar\psi}^j_+
\partial_{\xi^-}{\bar \phi}^k  
- F^i{\bar\psi}^j_-{\bar\psi}^k_+ )
{\tilde \Gamma}_{i\bar j \bar k}
+ (\psi^i_-{\bar\psi}^k_- 
\psi^j_+{\bar\psi}^l_+){\tilde {\mathcal K}}_{ij\bar k \bar l}
\,\Bigr].
\eea 
To simplify the action further, let us now look at the equation of
motion of the auxiliary fields.
Due to the various powers of the auxiliary field $F$ appearing
in the above action, equation of motion of ${\bar F}$ cannot
be derived directly. The equation of motion of $F$ at first
order in $(\det\,C)$ can however
be shown to be\footnote{Solutions
to auxiliary fields in four dimensions have also been 
discussed in~\cite{Inami:2004sq,Azorkina:2005mx}}
(after renaming some dummy indices):
\be \label{NewFeq}
{\tilde g}_{i{\bar j}}\, F^i + \psi^{k}_- \psi^{l}_+ \,  
{\tilde \Gamma}_{kl\bar j}  = 0.
\ee
This is infact quite similar to the solution for $F$ to lowest order
in $(\det\,C)$ given in eqn. (\ref{F}); but not quite the same, as
various geometric quantities are redefined. If one naively attempts to
check whether the lowest order solution can still work, i.e., by 
substituting the
result of eqn. (\ref{F}) in  (\ref{NewFeq}), one finds a non-zero piece of
the kind $\frac{1}{6}(\det\,C) F^{p}F^{q} \psi^{l}_- \psi^{m}_+ g_{i{\bar j}} 
\partial_p  \partial_q  \Gamma^i_{lm} $. 

Thus, in a nutshell, we have found an exact solution to the equations
of motion of  auxiliary field $F$, in terms of new
geometric quantities. Later we will argue that eqn. (\ref{NewFeq}) is in
fact an all order solution for $F$. 

In order to simplify the action further, terms bilinear in the 
bosonic fields have to be understood. For this, we recollect
that the classical action for non(anti)commutative sigma model 
can be divided in to two parts $I = I_o + I_c$, where $I_o$ and 
$I_c$ are the C-independent and the C-dependent parts respectively. 
$I_o$ is neatly
summarized in eqn. (\ref{Io}) and $I_c$ can deduced from eqn. (\ref{I}).
We note that  $I_o$ and $I_c$ can be shown to be independenly invariant under 
${\cal N} = 1/2$ supersymmetry of the theory~\cite{Chandrasekhar:2003uq},
and can be dealt seperately. We now use this fact to 
derive the equation of motion of auxiliary field $F$ for the C-dependent
and C-independent parts seperately, from eqn. (\ref{NewFeq}).
Collecting terms proportional to zeroth
and first order in $(\det\,C)$ on both sides of eqn. (\ref{NewFeq}), 
we have respectively:
\bea \label{Fnoc}
g_{i{\bar j}}\, F^i &=& -  \psi^{k}_- \psi^{l}_+ \,  
\Gamma_{kl\bar j} ,\\ \label{Fc1}
\partial_p \partial_q g_{i{\bar j}}\, F^i &=& -  \psi^{k}_- \psi^{l}_+ \,  
\partial_p \partial_q \Gamma_{kl\bar j} .
\eea
Thus, using eqn. (\ref{Fc1}) in eqn. (\ref{I0new}), the ${\cal N}=2$ 
non(anti)commutative $\sigma$-model action takes the standard form:
\bea \label{Ifinal}
I &=& \int d^2x \,\Bigl[\,
(\frac{1}{2}\partial_{\xi^-}\phi^i
\partial_{\zeta^-}{\bar \phi}^j
+\frac{1}{2}\partial_{\zeta^-}\phi^i
\partial_{\xi^-}{\bar \phi}^j
+i\psi^i_-\partial_{\zeta^-}{\bar\psi}^j_-
+ i\psi^i_+\partial_{\xi^-}
{\bar\psi}^j_+ + F^i{\bar F}^j )
{\tilde g}_{i\bar j} 
+ (\psi^i_-
\psi^j_+{\bar F}^k)  \xx
&\times& {\tilde \Gamma}_{ij\bar k} 
+ (i \psi^i_-{\bar\psi}^j_-\partial_{\zeta^-}{\bar \phi}^k
+ i \psi^i_+{\bar\psi}^j_+
\partial_{\xi^-}{\bar \phi}^k  
- F^i{\bar\psi}^j_-{\bar\psi}^k_+ )
{\tilde \Gamma}_{i\bar j \bar k}
+ (\psi^i_-{\bar\psi}^k_- 
\psi^j_+{\bar\psi}^l_+){\tilde {\mathcal K}}_{ij\bar k \bar l}
\,\Bigr],
\eea 
The fact that the non(anti)commutative sigma model action in 
eqn. (\ref{Ifinal}) is of the same form as the $C=0$ $\sigma$-model action 
is really intriguing. Notice however, that all the geometric quantities 
are redefined and contain $C$-dependent terms explicitly. For instance, 
the new metric is related to the old metric
by the addition of higher derivative $C$-dependent terms,
as seen in  eqn. (\ref{NewgG}). Thus, in the process of simplyfying the 
sigma model action, terms ensuing from the $C$-deformation on the worldsheet
have been transfered to the target space geometry. 

To be precise, in the action (\ref{Ifinal}), the redefined target space 
metric apperas fuzzy. This is rather counter intuitive, as only
the fermionic coordinates on the world-sheet
are deformed by the relations in eqn. (\ref{Cdeformation}). 
If the deformation was introduced for
the bosonic coordinates, then the bosonic components of the
chiral superfields will have to be multiplied using a new star
product, which have the effect of making the target space geometry fuzzy.
However, in the present case, the bosonic fields which act
as target space coordinates do not get affected by the 
deformation (\ref{Cdeformation}).  In the following,
we see whether going over to appropriate normal coordinates can 
undo the smearing of the target space geometry due to the deformation
terms.

Although, the analysis so far was for the sigma model action to
first order in $(\det\,C)$, it is straightforward to generalize the
results to the full action. The partial integrations pointed out in
eqn. (\ref{partial}) can be carried over in a similar way. However,
there will still be one piece remaining as in eqn. (\ref{I0new}).
In general, the bosonic part of the action will look as follows:
\bea
&& \int d^2x \,\Bigl[\,
(\frac{1}{2}\partial_{\xi^-}\phi^i
\partial_{\zeta^-}{\bar \phi}^j
+\frac{1}{2}\partial_{\zeta^-}\phi^i
\partial_{\xi^-}{\bar \phi}^j )\, (
g_{i\bar j} + \sum_{n=1}^{\infty}(\det\,C)^n\,(-1)^n 
F^{i_1}F^{i_2}\cdots F^{2n-1} \xx
&\times & \{ 
\frac{1}{(2n)}\,F^{2n}{\mathcal K}_{i_1\cdots i_{2n} i\bar j } 
+ \frac{1}{(2n+1)}\, \psi^{2n}_- \psi^{2n+1}_+ \,
{\mathcal K}_{i_1\cdots i_{2n+1} i\bar j } 
\} \, ) \,\Bigr]
\eea
To get the bosonic part in the standard form, one has to derive the 
equation of motion of auxiliary fields for the all order action 
in eqn. (\ref{I}). 

We claim that the auxiliary field $F$ still satisfies the same 
constraint equation as given in eqn. (\ref{F}). However, various
geometric quantities such as metric and christoffels will now 
contain many higher derivative terms in the deformation parameter $C$.
One can show this explicitly and we argue it to be so, by looking at
the following term in the all order action in eqn. (\ref{I}):
\be
\sum_{n=0}^{\infty} (\det\,C)^n (-1)^n
\:\int d^2x \: \frac {F^{i_1}F^{i_2}\cdots F^{i_{2n}}}{(2n+1)!} 
\Bigl(\, i \psi^{i_{2n+1}}_- \partial_{\zeta^-}{\bar\psi}^j_-  
+ i \psi^{i_{2n+1}}_+ \partial_{\xi^-} {\bar \psi}_+ 
+  F^{i_{2n+1}}{\bar F}^j\,\Bigr)
\partial_{i_1\cdots i_{2n}} \, g_{i_{2n+1} \bar j} .
\ee
From the above equation, one can first deduce how the various 
geometric quantities get redefined. For example, the metric picks
up many new $C$-dependent terms and takes the general form:
\be \label{metricNew}
{\tilde g}_{i{\bar j}}\, = g_{i{\bar j}} + 
\sum_{n=1}^{\infty}\, \frac{(-1)^n}{(2n+1)!}\,(\det\,C)^n  
F^{i_1}F^{i_2}\cdots F^{i_{2n}} 
{g}_{i \bar j,i_1\cdots i_{2n}} , 
\ee
where, as mentioned before, the subscripts of metric $g$ after
the comma, indicate derivatives with respect to 
the corresponding chiral superfield, evaluated at $\Phi = \phi$. One
can check the validity of the general form given in eqn. (\ref{metricNew}),
by analyzing other terms in the action (\ref{I}). The result in 
eqn. (\ref{NewgG}) can be obtained as a special case of $n=1$ from 
eqn. (\ref{metricNew}).

To summarize, the all order action in eqn. (\ref{I}) can be rewritten 
as in eqn. (\ref{I0new}) and the auxiliary field equations take the 
general form as in eqn. (\ref{NewFeq}). To rearrange the bosonic part of
the action, it is possible to derive relations similar to 
the ones given in eqns. (\ref{Fnoc}) and  (\ref{Fc1}), at every order
in $(\det\,C)$. Hence, at order $(\det\,C)^n$, $F$ satisfes the following 
constraint equation:
\be
 F^i \,\, \partial_{i_1\cdots i_{2n}} \,g_{i{\bar j}}\,
~=~ -  \psi^{k}_- \psi^{l}_+ \,\,  
 \partial_{i_1\cdots i_{2n}} \, \Gamma_{kl\bar j} \, .
\ee

Thus, by redefining certain geometric quantities the complicated looking 
action (\ref{I}) with infinite terms reduces to an extremly 
simple form and can be written succintly as in eqn. (\ref{Ifinal}). 
Since, the terms dependent on $(\det\,C)$ are
not present in the action explicitly, one can eliminate the auxiliary 
fields $F$ and ${\bar F}$ by their standard equations of motion. 
However, to show that the classical action for non(anti)commutative 
sigma model takes the final form as given in eqn. (\ref{Io}),
one needs to derive the inverse of new metric given in 
eqn. (\ref{metricNew}). Further, with the new metric, one also has to
show that ${\tilde \Gamma}^i_{kl} = {\tilde g}^{i{\bar j}}\, 
{\tilde \Gamma}_{kl\bar j} $. These issues are important in
understanding the Ricci-flatness conditions of the non(anti)commutative
sigma models.

%From eqn. (\ref{metricNew}),
%one notes that there is no direct way to guess the inverse of new metric.
%For instance, we can start with the assumption that an inverse 
%${\tilde g}^{i{\bar j}} $ exists and that 
%${\tilde g}_{i{\bar k}} \: {\tilde g}^{i{\bar j}} 
%= \delta^{\bar j}_{\bar k}$. 

Now, in terms of new coordinates, the action takes a form manifestly
invariant under general coordinate transformations. However, the
covariance of the action has to be explicitly shown, as the 
transformation properties of the new geometric quantities as given in
eqn. (\ref{metricNew}) are not clear. Since the NAC sigma model 
action takes a standard form, it can also be shown to be invariant
under the standard ${\cal N}=2$ supersymmetry transformations. This 
signifies the presence of enhanced supersymmetry in the theory and 
should be of interest\footnote{I wish to thank R. Gopakumar
for pointing this to me.}. Thus, in some sense, we have recovered
the standard ${\cal N}=2$ sigma model action on non(anti)commutative
superspace\footnote{For other discussions on this point 
see~\cite{REY1,TERA,Sako:2004at}.}, albeit with redefined 
geometric quantities. 

Let us try to understand eqn. (\ref{metricNew}) more carefully. One
notices that there are various powers of $(\det\,C\, F^2)$, which 
can be written in terms of superfields as follows:
\be \label{cf}
\det\,C\, F^2 = \det\,C\, (Q_+\,Q_-\, L)(Q_+\,Q_-\, L).
\ee
Here $(Q_+\,Q_-\, L)$ by itself is a new superfield and can appear 
in the action, since we only retain ${\cal N}=1/2$ 
supersymmetry~\cite{TERA}. Notice that we could have also 
chosen $\int d^2\theta \, L(Q_+\,Q_-\, L)$ for the superfield form.
However at higher orders, this form does not reproduce higher powers
of $(\det\,C) F^2$. Further, due to integration over half of superspace,
it is an $F$-type term and can only arise from the superpotential in
two dimensions. In this work we do not consider superpotential terms 
in the action. However, the resummation also works for the
superpotential terms, as has been discussed in~\cite{Alvarez-Gaume:2005pj}.

One should keep in mind that on a non(anti)commutative superspace, $F$ and 
$D$ terms are indistinguishable~\cite{REY1,TERA}, as new superfields 
of the kind ${\bar \theta}{\bar \theta}R$ etc., can be formed. Using this, 
one can change a $D$-term,
e.g., $\int d^2\theta  d^2{\bar \theta} L*({\bar \theta}{\bar \theta}R)$
to an $F$-term $\int d^2\theta \,L*R$. This should be compared with the case
in four dimensions, where the natural combination is 
$(\det\,C\, F^3)$ and one adds an extra factor of 
$L$ in eqn. (\ref{cf}). As a result, one requires an additional
superspace integral over half of the superspace which gives rise to
an $F$-type term and the NAC action can be written without 
using start product at the classical level. But, at the quantum level 
there are new terms which do not have classical counterparts~\cite{REY1,TERA}. 
New terms in the effective action can arise from the new vertex and 
significantly modify the renormalization properties of the theory at the 
quantum level~\cite{GRIS,Britto:2003aj}. It should 
be interesting to explore these effects in the two dimensions as well.

Though, we have been
working with the component form of the action, for rest of the analysis
it will be convenient to work with the superfield form of the action.
Using relation (\ref{cf}) we can write the redefined 
metric (\ref{metricNew}) as:
\be \label{gL}
{\tilde g}_{i{\bar j}}\, = g_{i{\bar j}} + 
\sum_{n=1}^{\infty}\,   \frac{(-1)^n}{(2n+1)!}\,(\det\,C)^n 
\, [Q_+\,Q_-\, L]^{2n}\,
{g}_{i \bar j,i_1\cdots i_{2n}} 
\ee
Further, one can rewrite the non(anti)commutative sigma model action
given in eqn. (\ref{Ifinal}), in terms of the superfields  
as follows (after some partial integrations):
\be \label{Action}
I_0 =~ \int d^2x \,d^2\theta \: \Bigl[\:
%R \,\frac{\partial{\mathcal K}}{\partial\bar{S}}
%\,+\,\frac{1}{2!}\,R_*^2\,
%\frac{\partial^2{\mathcal K}}{\partial \bar{S}^2}\,+\, 
\frac{1}{2!} \,\,L^i*R^j\, {\tilde g}_{i{\bar j}}\,
+\,\frac{1}{3!}\,L^i*L^j*R^k\,  {\tilde \Gamma}_{ij\bar k} 
\,+\,  \frac{1}{3!}\,L^i*R^j*R^k \, {\tilde \Gamma}_{i\bar j\bar k}
\,+\, \cdots \:\Bigr].
\ee
To explore the relation between ${\tilde g}$ and $g$, 
let us consider the first term in the action (\ref{Action}). In other
words, we multiply both sides of eqn. (\ref{gL}) by $L^{i}*R^{j}$ and
explicitly write down the first few terms in eqn. (\ref{gL}) as:
\bea \label{gL2}
{\tilde g}_{i{\bar j}}\,L^{i}*R^{j}\, &=& \,g_{i{\bar j}}  \,
L^{i}*R^{j}\, +   \frac{1}{3!} \, (Q_+Q_- L)^{i_1}*(Q_+Q_- L)^{i_2}*
L^{i}*R^{j}\,\partial_{i_2}\,{\Gamma}_{i i_1\bar j} \xx
&+& \frac{1}{5!} \, (Q_+Q_- L)^{i_1}*(Q_+Q_- L)^{i_2}*
(Q_+Q_- L)^{i_3}*(Q_+Q_- L)^{i_4}* \xx
&\times&L^{i}*R^{j}\,\partial_{i_2}\partial_{i_3}\partial_{i_4}\,
{\Gamma}_{i i_1\bar j} \,+\, \cdots .
\eea
Renaming some dummy indices, we finally arrive at:
\bea \label{gL3}
{\tilde g}_{i{\bar j}}\,L^{i}*R^{j}\, &=& \,g_{m{\bar l}}  \,
[ L^{m} +   \frac{1}{3!} \, (Q_+Q_- L)^{i_1}*(Q_+Q_- L)^{i_2}*L^{i}\,
\partial_{i_2}\,{\Gamma}_{i i_1\bar j} \, g^{m{\bar j}} + \frac{1}{5!} \, 
(Q_+Q_- L)^{i_1} \xx
&\times& *(Q_+Q_- L)^{i_2}* (Q_+Q_- L)^{i_3}*(Q_+Q_- L)^{i_4}*
L^{i}\partial_{i_2}\partial_{i_3}\partial_{i_4}\,
{\Gamma}_{i i_1\bar j} \, g^{m{\bar j}}  +\,\cdots ]\,*R^{l}\, , \xx
\label{gL4}
&=& \,g_{m{\bar l}}  \,[   L^{m} 
+ \sum_{n=1}^{\infty}\,  \frac{1}{(2n+1)!} \,
g^{m{\bar j}}\,\partial_{i_3\cdots i_{2n+1}}
\Gamma_{i_1i_2\bar j}\,(Q_+Q_- L)^{2n}\,L^{2n+1}]\,*R^{l}\, ,
\eea
We rewrite the the quantity in square brackets in eqn. (\ref{gL4}) as:
\be \label{KNC}
\pi^m =   L^{m} 
+ \sum_{n=1}^{\infty}\,  \frac{1}{(2n+1)!} \,
g^{m{\bar j}}\,\partial_{i_3\cdots i_{2n+1}}
{\mathcal K}_{i_1i_2\bar j}\,(Q_+Q_- L)^{2n}\,L^{2n+1},
\ee
and note that there are no star products in the definition of  $\pi$.
The new coordinates $\pi$ resemble the K\"{a}hler normal coordinates
which were introduced in ref.~\cite{Higashijima:2000wz}. These normal
coordinates have the advantage that they transform as holomorphic 
tangent vectors at the origin. However, in our case it remains
to be seen whether  $\pi$'s can be taken as new normal coordinates. 

In ref.~\cite{Higashijima:2000wz}, using mathematical induction 
it was explicitly shown that 
$\pi$'s transform as holomorphic tangent vectors on the target 
K\"{a}hler manifold. However, for the present case the proof
does not go through, due to the presence of auxliary fields in the
new coordinates. Notice, we have many factors of $(Q_+Q_- L)= F$ appearing
in eqn. (\ref{KNC}). We know that, under a general coordinate 
transformation of the target space manifold given as, 
$\Phi'(x,\theta,{\bar\theta}) = f (\Phi)$, the auxiliary field $F$ 
transforms as:
\be
F'^i(x) = \frac{\partial f^i(\phi(x))}{\partial \phi^j} F^j -
\frac{1}{2} \frac{\partial^2 f(\phi(x))}{\partial \phi^j\partial \phi^k}
\psi_-^j\psi_+^k
\ee
Although it is still true that in eqn. (\ref{KNC}), the generalized connection 
${\mathcal K}_{i_1i_2 \cdots i_{2n+1}\bar j}$ transforms as in the standard
case~\cite{Higashijima:2000wz}, the auxiliary fields transformations
have an inhomogenious piece in fermions. In the proof outlined 
in~\cite{Higashijima:2000wz}, instead of $(Q_+Q_- L)$ one has various
powers of $L$, which transform 
smoothly (see Appendix C. of~\cite{Higashijima:2000wz}). 
One can explicitly show that
$\pi$'s do not transform as holomorphic tangent vectors on the target
manifold.

The above analysis suggests that $\pi$'s cannot be taken as new normal
coordinates. Also, it is not possible to remove the deformation terms in the 
eqn. (\ref{metricNew}), by doing any kind of holomorphic coordinate
transformation of the target manifold. This asserts that, the deformation of 
target space due to terms dependent on various powers of auxliary 
field $F$ and $(\det\,C)$, is a generic feature of 
non(anti)commutative sigma models.

The redefinition of the target space metric can also be understood as
the redefinition of the K\"{a}hler potential itself, as discussed 
in~\cite{Alvarez-Gaume:2005pj}. All the  $C$-dependent terms appearing 
in the redefined geometric quantities are non-covariant. The advantage 
of working with 
K\"{a}hler normal coordinates is that all the terms appearing in 
the expansion are guaranteed to be covariant. Thus, working with 
proper K\"{a}hler normal coordinates and looking at 
the expansions of various geometric quantities to a few orders, might 
give a hint of the smearing of the target 
space~\cite{Higashijima:2000wz,Chandrasekhar:2004ti}.

Due to the redefinition of various geometric quantities, the symmetry 
between the holomorphic and anti-holomophic terms in the action
is formally restored. This symmetry was previously absent, as various
powers of $(\det\,C)$ appeared with powers of $F$, but there 
were no corresponding ${\bar F}$ pieces in the action (\ref{I}).   
Since, the non(anti)commtuative sigma model action takes a standard form,
it should be possible to extend the results of $C=0$ sigma models, 
such as quantum renormalizability and conformal invariance to the
present case as well.

To take an example, for the standard sigma 
models at the quantum level, the one-loop $\beta$-function goes 
as $ k\,\Tr \ln g $, where
$k$ is a constant. Now, for the present case, there are no additional
vertices at the quantum level, unlike the four dimensional case where 
$(\det\, C F^3)$ leads to many new features. Further, since 
the $\beta$-function calculation only depends on the metric, it is tempting
to identify the beta function of the non(anti)commutative sigma models
with that of the standard sigma models, where $g$ is replaced by $\tilde g$.
It should be interesting to find out how the Ricci flat conditions
look like, in terms of new geometric quantities. Especially, how do
the terms dependent on $C$ modify the conformal invariance conditions 
of the non(anti)commutative sigma models. However, to make concrete 
progress, the smearing of the target space quantities due the terms 
dependent on $C$ and the transformation properties of $\tilde g$ have
to be better understood. It is also essential to derive various
geometric quantities starting from $\tilde g$ and show that they 
are globally well defined on the target
manifold~\cite{Nemeschansky:1986yx}. Further, one
should note that the sigma models studied here are inherently of Euclidean
nature and it is interesting to understand how to surpass the conditions
in~\cite{Klemm:2001yu} to define the same on minkowski spaces.

\begin{center}
{\bf Acknowledgments}
\end{center}

I wish to thank A. Kumar and A. Sen for useful suggestions. I thank
the organizers of International Workshop on String 
theory (ISM04), Khajuraho, for giving me an opportunity to present 
this work. I am grateful to the organizers
of 9th APCTP/KIAS winter school on string theory, Seoul, for support and
warm hospitality during the final stages of this work.

%\appendix
%\section*{{Appendix A.}}
%\renewcommand{\theequation}{A-\arabic{equation}}
%\setcounter{equation}{0}

%{\cal D}
%eqn. (\ref{})
%%%%%%%%%%%%%%%%%%%%%%%%%%%%%%%%%%%%%%%%%%%%%%%%%%%%%%%%%%%%%%%%%%%%%%%%
%                       REFERENCES                                     %
%%%%%%%%%%%%%%%%%%%%%%%%%%%%%%%%%%%%%%%%%%%%%%%%%%%%%%%%%%%%%%%%%%%%%%%%
%\newpage


\begin{thebibliography}{99}

%\cite{Zumino:1979et}
\bibitem{Zumino:1979et}
B.~Zumino,
%%``{\em Supersymmetry And Kahler Manifolds},''
Phys.\ Lett.\ B {\bf 87}, 203 (1979).

%\cite{Friedan:1980jf}
\bibitem{Friedan:1980jf}
D.~Friedan,
%%``{\em Nonlinear Models In Two Epsilon Dimensions},''
Phys.\ Rev.\ Lett.\  {\bf 45}, 1057 (1980).

%\cite{Alvarez-Gaume:1980dk}
\bibitem{Alvarez-Gaume:1980dk}
L.~Alvarez-Gaume and D.~Z.~Freedman,
%%``{\em Kahler Geometry And The Renormalization Of
%%Supersymmetric Sigma Models},''
Phys.\ Rev.\ D {\bf 22}, 846 (1980) ; 
%%``{\em Geometrical Structure And Ultraviolet Finiteness 
%%In The Supersymmetric Sigma Model},''
Commun.\ Math.\ Phys.\  {\bf 80}, 443 (1981).


%\cite{Alvarez-Gaume:hn}
\bibitem{Alvarez-Gaume:hn}
L.~Alvarez-Gaume, D.~Z.~Freedman and S.~Mukhi,
%%``{\em The Background Field Method And The Ultraviolet 
%%Structure Of The Supersymmetric Nonlinear Sigma Model},''
Annals Phys.\  {\bf 134}, 85 (1981).

%\cite{Seiberg:1999vs}
\bibitem{Seiberg:1999vs}
N.~Seiberg and E.~Witten,
%%``{\em String theory and noncommutative geometry,}''
JHEP {\bf 9909}, 032 (1999)
[hep-th/9908142]. 
%%CITATION = HEP-TH 9908142;%%

%\cite{Girotti:2001gs}
\bibitem{Girotti:2001gs}
  H.~O.~Girotti, M.~Gomes, V.~O.~Rivelles and A.~J.~da Silva,
  %``The noncommutative supersymmetric nonlinear sigma model,''
  Int.\ J.\ Mod.\ Phys.\ A {\bf 17}, 1503 (2002)
  [hep-th/0102101].
  %%CITATION = HEP-TH 0102101;%%


%\cite{Ooguri:2003qp}
\bibitem{Ooguri:2003qp}
H.~Ooguri and C.~Vafa,
%%``{\em The C-deformation of gluino and non-planar diagrams},''
Adv.\ Theor.\ Math.\ Phys.\  {\bf 7}, 53 (2003),
[hep-th/0302109] ; J.~de Boer, P.~A.~Grassi and P.~van Nieuwenhuizen,
%%``{\em Non-commutative superspace from string theory},''
Phys.\ Lett.\ B {\bf 574}, 98 (2003),
[hep-th/0302078].
%%CITATION = HEP-TH 0302109;%%


%\cite{Seiberg:2003yz}
\bibitem{Seiberg:2003yz}
N.~Seiberg,
%% ``{\em Noncommutative superspace, N = 1/2 supersymmetry, 
%%field theory and  string theory,}''
JHEP {\bf 0306}, 010 (2003)
[hep-th/0305248].


%\cite{Berkovits:2003kj}
\bibitem{Berkovits:2003kj}
  N.~Berkovits and N.~Seiberg,
  %``Superstrings in graviphoton background and N = 1/2 + 3/2 supersymmetry,''
  JHEP {\bf 0307}, 010 (2003)
  [hep-th/0306226].
  %%CITATION = HEP-TH 0306226;%%

%\cite{Klemm:2001yu}
\bibitem{Klemm:2001yu}
  D.~Klemm, S.~Penati and L.~Tamassia,
  %``Non(anti)commutative superspace,''
  Class.\ Quant.\ Grav.\  {\bf 20}, 2905 (2003)
  [hep-th/0104190].
  %%CITATION = HEP-TH 0104190;%%


%\cite{Ferrara:2000mm}
\bibitem{Ferrara:2000mm}
  S.~Ferrara and M.~A.~Lledo,
  %``Some aspects of deformations of supersymmetric field theories,''
  JHEP {\bf 0005}, 008 (2000)
  [hep-th/0002084];
  %%CITATION = HEP-TH 0002084;%%
%\cite{Ferrara:2003xy}
%\bibitem{Ferrara:2003xy}
  S.~Ferrara, M.~A.~Lledo and O.~Macia,
  %``Supersymmetry in noncommutative superspaces,''
  JHEP {\bf 0309}, 068 (2003)
  [hep-th/0307039].
  %%CITATION = HEP-TH 0307039;%%

%\cite{Ivanov:2003te}
\bibitem{Ivanov:2003te}
  E.~Ivanov, O.~Lechtenfeld and B.~Zupnik,
  %``Nilpotent deformations of N = 2 superspace,''
  JHEP {\bf 0402}, 012 (2004)
  [hep-th/0308012].
  %%CITATION = HEP-TH 0308012;%%


%\cite{Chandrasekhar:2003uq}
\bibitem{Chandrasekhar:2003uq}
B.~Chandrasekhar and A.~Kumar,
%%``{\em D = 2, N = 2 supersymmetric theories on 
%%non(anti)commutative superspace,}''
JHEP {\bf 0403}, 013 (2004)
[hep-th/0310137].
%%CITATION = HEP-TH 0310137;%%

%\cite{Chandrasekhar:2004ti}
\bibitem{Chandrasekhar:2004ti}
  B.~Chandrasekhar,
  %``D = 2, N = 2 supersymmetric sigma models on non(anti)commutative
  %superspace,''
  Phys.\ Rev.\ D {\bf 70}, 125003 (2004)
  [hep-th/0408184].
  %%CITATION = HEP-TH 0408184;%%

%\cite{Alvarez-Gaume:2005pj}
\bibitem{Alvarez-Gaume:2005pj}
  L.~Alvarez-Gaume and M.~A.~Vazquez-Mozo,
  %``On nonanticommutative N=2 sigma-models in two dimensions,''
  [hep-th/0503016].
  %%CITATION = HEP-TH 0503016;%%



\bibitem{REY1}
R. Britto, B. Feng, and S.J. Rey, 
%%{\em Deformed Superspace, N=1/2
%%Supersymmetry and (Non)Renormalization Theorems}, 
JHEP 0307 (2003)067, [hep-th/0306215].



%\cite{Lukierski:1986jw}
\bibitem{Lukierski:1986jw}
  J.~Lukierski and W.~J.~Zakrzewski,
  %``Euclidean Supersymmetrization Of Instantons And Selfdual Monopoles,''
  Phys.\ Lett.\ B {\bf 189}, 99 (1987).
  %%CITATION = PHLTA,B189,99;%%

\bibitem{TERA} S. Terashima, and J. Yee, 
%{\em Comments on Noncommutative
%Superspace}, 
JHEP {\bf 0312}, 053 (2003),
[hep-th/0306237].


%\cite{Inami:2004sq}
\bibitem{Inami:2004sq}
  T.~Inami and H.~Nakajima,
  %``Supersymmetric CP(N) sigma model on noncommutative superspace,''
  Prog.\ Theor.\ Phys.\  {\bf 111}, 961 (2004)
  [hep-th/0402137].
  %%CITATION = HEP-TH 0402137;%%

%\cite{Azorkina:2005mx}
\bibitem{Azorkina:2005mx}
  O.~D.~Azorkina, A.~T.~Banin, I.~L.~Buchbinder and N.~G.~Pletnev,
  %``Generic chiral superfield model on nonanticommutative 
%N = 1/2 superspace,''
  [hep-th/0502008].
  %%CITATION = HEP-TH 0502008;%%




\bibitem{GRIS}
M.~T.~Grisaru, S.~Penati, and A.~Romagnoni,
%%{\em Two-loop Renormalization for Nonanticommutative N=1/2 Supersymmetric
%%WZ Model}, 
JHEP 0308 (2003) 003, [hep-th/0307099] ; R.~Britto, and B.~Feng, 
%%{\em N=1/2 Wess-Zumino model is renormalizable},
Phys.\ Rev.\ Lett.\  {\bf 91}, 201601 (2003), [hep-th/0307165]; 
A.~Romagnoni, 
%%{\em Renormalizability of N=1/2 Wess-Zumino model in
%superspace}, 
JHEP {\bf 0310}, 016 (2003),
[hep-th/0307209].

%\cite{Britto:2003aj}
\bibitem{Britto:2003aj}
  R.~Britto, B.~Feng and S.~J.~Rey,
  %``Non(anti)commutative superspace, UV/IR mixing and open Wilson lines,''
  JHEP {\bf 0308}, 001 (2003)
  [hep-th/0307091];
  %%CITATION = HEP-TH 0307091;%%
%\cite{Lunin:2003bm}
%\bibitem{Lunin:2003bm}
  O.~Lunin and S.~J.~Rey,
  %``Renormalizability of non(anti)commutative gauge theories with N = 1/2
  %supersymmetry,''
  JHEP {\bf 0309} (2003) 045
  [hep-th/0307275].
  %%CITATION = HEP-TH 0307275;%%



%\cite{Sako:2004at}
\bibitem{Sako:2004at}
  A.~Sako and T.~Suzuki,
  %``Recovery of full N = 1 supersymmetry in non(anti-)commutative superspace,''
  JHEP {\bf 0411}, 010 (2004)
  [hep-th/0408226].
  %%CITATION = HEP-TH 0408226;%%



%\cite{Higashijima:2000wz}
\bibitem{Higashijima:2000wz}
K.~Higashijima and M.~Nitta,
%``Kaehler normal coordinate expansion in supersymmetric theories,''
Prog.\ Theor.\ Phys.\  {\bf 105}, 243 (2001)
[hep-th/0006027];
%%CITATION = HEP-TH 0006027;%%
%\cite{Higashijima:2002fq}
%%\bibitem{Higashijima:2002fq}
K.~Higashijima, E.~Itou and M.~Nitta,
%``Normal coordinates in Kaehler manifolds and the background field  method,''
Prog.\ Theor.\ Phys.\  {\bf 108}, 185 (2002)
[hep-th/0203081].
%%CITATION = HEP-TH 0203081;%%

%\cite{Nemeschansky:1986yx}
\bibitem{Nemeschansky:1986yx}
D.~Nemeschansky and A.~Sen,
%%``{\em Conformal Invariance Of Supersymmetric Sigma Models On Calabi-Yau
%%Manifolds,}''
Phys.\ Lett.\ B {\bf 178}, 365 (1986).
%%CITATION = PHLTA,B178,365;%%








\end{thebibliography}
\end{document}